\documentclass[manuscript]{acmart}
\usepackage{subfig}

\AtBeginDocument{%
  \providecommand\BibTeX{{%
    \normalfont B\kern-0.5em{\scshape i\kern-0.25em b}\kern-0.8em\TeX}}}

\setcopyright{acmcopyright}
\copyrightyear{2021}
\acmYear{2021}
\acmDOI{10.1145/3460231.3474599}

\acmConference[RecSys '21]{RecSys '21:  ASDF}{September 03--05, 2021}{Amsterdam}
\acmBooktitle{Fifteenth ACM Conference on Recommender Systems (RecSys ’21),
  September 03--05, 2021, Amsterdam}
\acmPrice{15.00}
\acmISBN{978-1-4503-XXXX-X/18/06}

\begin{document}

\title{Online Learning for Recommendations at Grubhub}

\author{Alex Egg}
\email{eggie5@hey.com}

\renewcommand{\shortauthors}{Alex Egg}

\begin{abstract}
We propose a method to easily modify existing offline Recommender Systems to run online using Transfer Learning.  Online Learning for Recommender Systems has two main advantages: quality and scale. Like many Machine Learning algorithms in production if not regularly retrained will suffer from Concept Drift. A policy that is updated frequently online can adapt to drift faster than a batch system. This is especially true for user-interaction systems like recommenders where the underlying distribution can shift drastically to follow user behaviour. As a platform grows rapidly like Grubhub, the cost of running batch training jobs becomes material. A shift from stateless batch learning offline to stateful incremental learning online can recover, for example, at Grubhub, up to a 45x cost savings and a +20\% metrics increase. There are a few challenges to overcome with the transition to online stateful learning, namely convergence, non-stationary embeddings and off-policy evaluation, which we explore from our experiences running this system in production.
\end{abstract}

\ccsdesc[500]{Information systems~Personalization}

\keywords{Grubhub, Online Learning, Transfer Learning}

\maketitle

\section{Introduction}

Grubhub is an online network of localized marketplaces that connect diners and local restaurants for delivery or pickup meals. Grubhub runs a large-scale recommender system that surfaces relevant restaurants based on a user profile and real-time context.  

Recommendations drive 80\% of revenue at Grubhub, therefore the quality of recommender systems is very important as it has a material impact on business. Additionally, Grubhub has many millions of diners and restaurants and is growing fast, therefore efficient scaling of recommender systems in terms of time and money is important.

\subsection{Data Drift}

Machine Learning models perform best right after training. In production, models degrade quickly because of concept drift. In Recommender Systems this is often realized as item catalog updates, user interest/preference shifts or more often, current events such as the recent pandemic or regional holidays such as Cinco de Mayo. 

To demonstrate this phenomena, we launched 2 experiments, one had no retraining at all and the other had daily retraining. Results are shown below in table \ref{table:1}, as relative increase in Purchase Through Rate (PTR) over the baseline with no retraining.

\begin{table}[h!]
\centering
\begin{tabular}{ c c }
 Retraining Method & \% PTR Increase  \\ 
 \hline
 No Retraining (Baseline) & 0.0  \\  
 Weekly Retraining & +2.5   \\
 Daily Retraining & +20.3    
\end{tabular}
\caption{Concept Drift: Relative lift against baseline system for daily and weekly retraining. These results show the importance of a fast response to drift.}
\label{table:1}
\end{table}

These results, in Table \ref{table:1}, suggest that at updates a daily cadence, if not more frequent, are required to use machine learning in production for our recommender use-case. This is caused by drift. This realization, that drift affects business metrics, motivates and necessitates not just model retraining, but frequent model retraining to allow our system to rapidly adapt to data distribution changes.

\subsection{Retraining}

One practical approach to periodic retraining is to move a fixed-size sliding window over a range of $n$ dates as new data becomes available. For example, one may choose to train on n=4 days of data, everyday. This is illustrated below in Figure \ref{fig:stateless}.

\begin{figure}[h]
    \centering
    \subfloat[\centering Stateless \label{fig:stateless} ] {\includegraphics[width=0.4\textwidth]{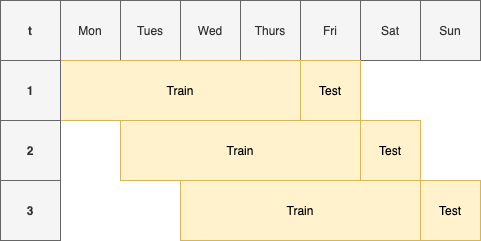}}%
    \qquad
    \subfloat[\centering Stateful \label{fig:stateful} ]{ \includegraphics[width=0.4\textwidth]{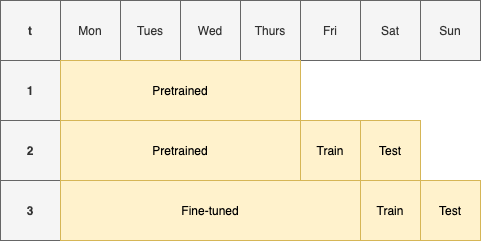}}%
\caption{\ref{fig:stateless} Example of daily Stateless updates with a 4-day sliding-window and next day hold-out for Cross Validation. \ref{fig:stateful} Example of daily incremental updates with state \& Bootstrapping. Notice the model is only updating on \textit{new} data, not the sliding-window in Fig \ref{fig:stateless}. The cadence can be 1-day, as in this example, or as small as a 10 minutes as in \cite{10.1145/3383313.3412214}}
\label{fig:stateless-stateful}
\end{figure}

In theory this may seem reasonable, however in practice, as many organizations are renting compute from the public cloud, this is \textit{expensive}, in terms of time and money, as $(n-1)$ days of data are redundantly computed every retraining session. If we could remove the redundant computations in our retraining routine, we could theoretically achieve a $nx$ time and cost savings. For the example in figure \ref{fig:stateless}, a $4x$ cost savings could be realized. Concretely, in early 2021, a p3.8xl on AWS costs \$12/hour. A representative retraining job at Grubhub takes about 3 hours to learn on 30 days of data. That comes out to approximately \$1080/month or 3x wasted compute. Adding state out our retraining system can help solve our cost issues by removing redundant computations over data.

\section{Stateful Updates}

Traditional supervision pipelines are stateless, meaning that in between rounds of retraining, state is not preserved. This is simpler from an ops perspective, however, it comes with the cost of computations over redundant data.

Out-of-core/incremental algorithms can be updated stochastically, meaning, they can be updated without having all the data at once. Parametric models, such as neural networks are one example, however, the family decision tree algorithms is not \cite{domingos2000mining}*. Most out-of-core models can be trivially migrated to be updated online. The key is to maintain state between training sessions which practically means serializing and deserializing before and after updates respectfully. This is illustrated in figure \ref{fig:stateful}, where the running example is adapted to be stateful. As you can see the model is updated incrementally only on new data.

\begin{figure}%
    \centering
    \subfloat[\centering Batch Updates\label{fig:batch}] {
        {
            \includegraphics[width=0.75\textwidth] {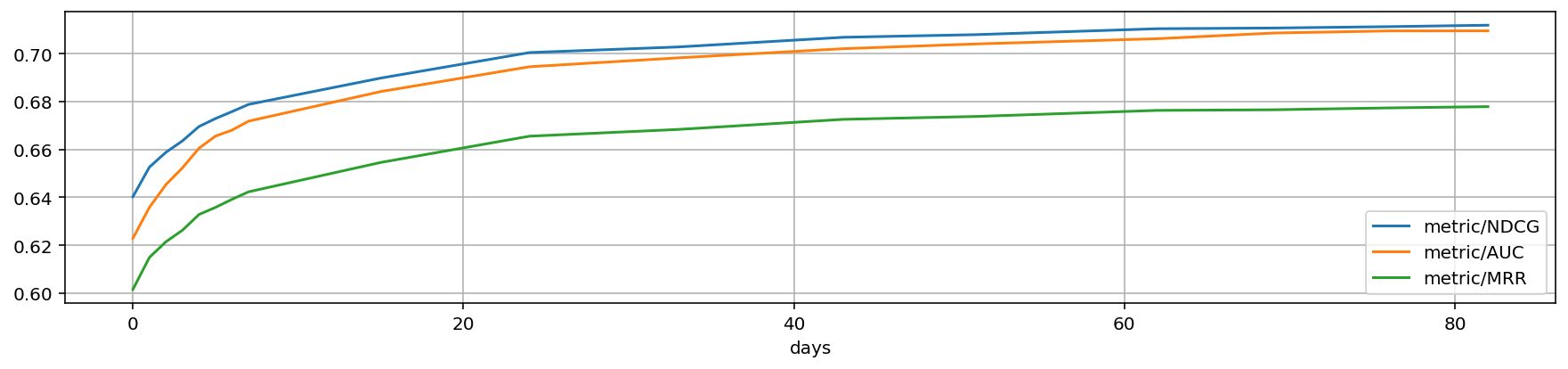} 
        }
    }%

    \subfloat[\centering Incremental Updates \label{fig:bonline}] {{\includegraphics[width=0.75\textwidth]{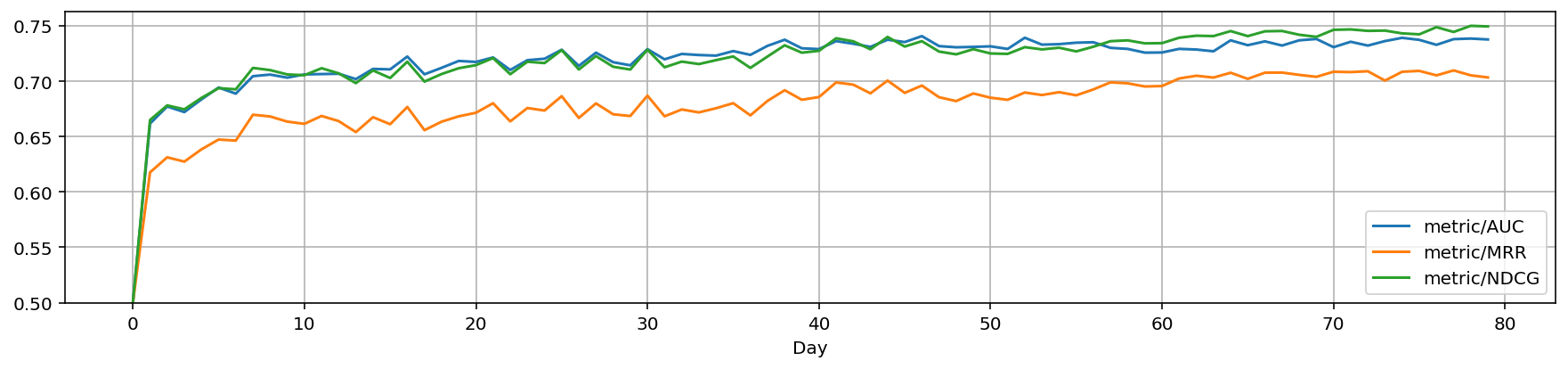} }}%
    \caption{\ref{fig:batch} Batch Cross Validation results on incrementally increasing training window from 1 to 80 days. \ref{fig:bonline} Incremental updates for 80 days. Converges faster than batch updates in Figure \ref{fig:batch}}%
    \label{fig:incremental}%
\end{figure}

To demonstrate this idea, we retrained daily two models both on 80 days of data, where one is stateless and one is stateful. In the minimum our hypothesis is that the incremental model will recover the batch model. Fig \ref{fig:incremental} show the incremental approach converges faster and to a higher maximum. In other words, the stateful approach does indeed recover the stateless approach, i.e. \textit{ Fig \ref{fig:batch} and \ref{fig:bonline} are identical as far as cross validation is concerned.}

\section{Off-Policy Evaluation \&  Pre-training}

An analogy can be drawn from the Computer Vision field where it is not uncommon to take a pre-trained imagenet model and then fine-tune it on your custom task/domain. Through this lens, \textit{retraining for drift can be viewed as a domain adaptation via transfer-learning}. In our case we are training offline until convergence and then fine-tuning online to rapidly adapt for drift.

In the online setting, the notion of hyper-parameter tuning, cross validation and even convergence become blurred. However, like mentioned above, we are only fine-tuning online and not necessarily learning from scratch online. Log data can be leveraged for pre-training which allows us to back-test the model offline using a simulator that walks through stochastic updates. After the model has converged, then we can release it for online updates.

\section{Embeddings in Online Learning}

Vocabularies that map categorical features to an integer index in an embedding table require advanced knowledge of all possible categories. This is unfeasible in the ecomm/product setting where products/restaurants come and go in real-time. One technique to map from from non-stationary categories to an integer index is a hash map.

A hash can map any given input into 1 of $n$ buckets (size) at the expense of collisions. The challenge is to tune $n$ with respect to an acceptable collision rate. Double Hashing \cite{10.1145/3383313.3412227} introduces two (n) independent hashing functions which reduces the collision rate exponentially, with only twice the memory consumption because a collision can only occur when both collide.

In order the quantify this trade-off we map the world-wide restaurant catalog to a range of sequential integers (buckets) using a hash function. We increase the buckets from 500k to 4 million and record the respective collision rates below in Fig \ref{fig:collisions} .

Figure \ref{fig:collisions} shows the collision rate converges to about 7\% at 3M buckets. Accordingly, we can be assured that with this encoding scheme we can represent most restaurants in the world and can adapt to new restaurants online, without full retraining. One sacrifice we make for using hashing is that there is no way to compute the inverse transform which can be a problem when trying to introspect which features are most important to a model.

\begin{figure}[h]
    \centering
\includegraphics[width=0.4\textwidth]{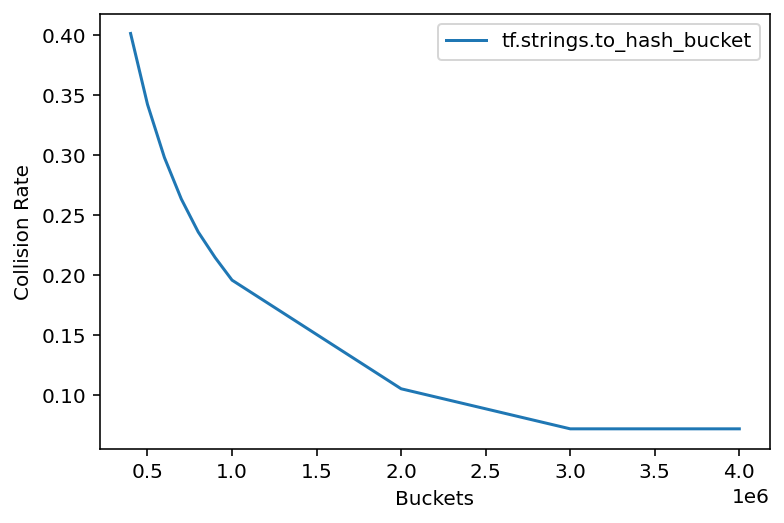}
\caption{Collision Rate vs Buckets.}
\label{fig:collisions}
\end{figure}

\section{Results}

\begin{figure}[h]
    \centering
\includegraphics[width=0.7\textwidth]{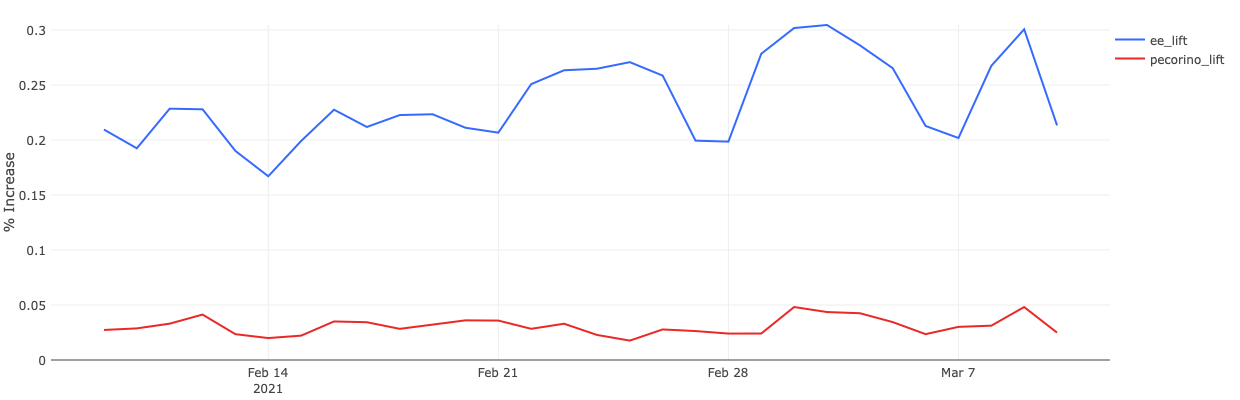}
\caption{Experiment showing relative increase in PTR for different update schemes.}
\label{fig:ab}
\end{figure}

Figure \ref{fig:ab} shows the AB Test results for the online migration. Results show a +20\% increase over the baseline PTR and additionally a 45x cost decrease. We attribute the results to faster drift response and decreased cloud usage respectively. Overall a realization of hundreds of thousands of dollars annually.

We also introduced a pre-training and fine-tuning technique as bridge between offline and online learning. Traditionally, the gap between the two paradigms has been technologically enormous, which effectively prevents participation except for the largest organizations with large resources \cite{chipwebsite}. This work has substantially shifted the bar between cost and rapid results by removing some of the traditional constraints limiting organizations to the offline paradigm.

\bibliographystyle{ACM-Reference-Format}
\bibliography{sample-manuscript}


\begin{thebibliography}{4}


\ifx \showCODEN    \undefined \def \showCODEN     #1{\unskip}     \fi
\ifx \showDOI      \undefined \def \showDOI       #1{#1}\fi
\ifx \showISBNx    \undefined \def \showISBNx     #1{\unskip}     \fi
\ifx \showISBNxiii \undefined \def \showISBNxiii  #1{\unskip}     \fi
\ifx \showISSN     \undefined \def \showISSN      #1{\unskip}     \fi
\ifx \showLCCN     \undefined \def \showLCCN      #1{\unskip}     \fi
\ifx \shownote     \undefined \def \shownote      #1{#1}          \fi
\ifx \showarticletitle \undefined \def \showarticletitle #1{#1}   \fi
\ifx \showURL      \undefined \def \showURL       {\relax}        \fi
\providecommand\bibfield[2]{#2}
\providecommand\bibinfo[2]{#2}
\providecommand\natexlab[1]{#1}
\providecommand\showeprint[2][]{arXiv:#2}

\bibitem[\protect\citeauthoryear{Domingos and Hulten}{Domingos and
  Hulten}{2000}]%
        {domingos2000mining}
\bibfield{author}{\bibinfo{person}{P Domingos} {and} \bibinfo{person}{G
  Hulten}.} \bibinfo{year}{2000}\natexlab{}.
\newblock \showarticletitle{Mining high-speed data streams.[In:] Proceedings of
  the sixth ACM SIGKDD International Conference on Knowledge Discovery and Data
  Mining}.
\newblock \bibinfo{journal}{\emph{Boston}}  \bibinfo{volume}{71}
  (\bibinfo{year}{2000}), \bibinfo{pages}{80}.
\newblock


\bibitem[\protect\citeauthoryear{Guo, Ktena, Myana, Huszar, Shi, Tejani,
  Kneier, and Das}{Guo et~al\mbox{.}}{2020}]%
        {10.1145/3383313.3412214}
\bibfield{author}{\bibinfo{person}{Dalin Guo}, \bibinfo{person}{Sofia~Ira
  Ktena}, \bibinfo{person}{Pranay~Kumar Myana}, \bibinfo{person}{Ferenc
  Huszar}, \bibinfo{person}{Wenzhe Shi}, \bibinfo{person}{Alykhan Tejani},
  \bibinfo{person}{Michael Kneier}, {and} \bibinfo{person}{Sourav Das}.}
  \bibinfo{year}{2020}\natexlab{}.
\newblock \showarticletitle{Deep Bayesian Bandits: Exploring in Online
  Personalized Recommendations}. In \bibinfo{booktitle}{\emph{Fourteenth ACM
  Conference on Recommender Systems}} (Virtual Event, Brazil)
  \emph{(\bibinfo{series}{RecSys '20})}. \bibinfo{publisher}{Association for
  Computing Machinery}, \bibinfo{address}{New York, NY, USA},
  \bibinfo{pages}{456–461}.
\newblock
\showISBNx{9781450375832}
\urldef\tempurl%
\url{https://doi.org/10.1145/3383313.3412214}
\showDOI{\tempurl}


\bibitem[\protect\citeauthoryear{Huyen}{Huyen}{[n.d.]}]%
        {chipwebsite}
\bibfield{author}{\bibinfo{person}{Chip Huyen}.}
  \bibinfo{year}{[n.d.]}\natexlab{}.
\newblock \bibinfo{booktitle}{\emph{Machine learning is going real-time}}.
\newblock
\urldef\tempurl%
\url{https://huyenchip.com/2020/12/27/real-time-machine-learning.html}
\showURL{%
\tempurl}


\bibitem[\protect\citeauthoryear{Zhang, Liu, Xie, Ktena, Tejani, Gupta, Myana,
  Dilipkumar, Paul, Ihara, Upadhyaya, Huszar, and Shi}{Zhang
  et~al\mbox{.}}{2020}]%
        {10.1145/3383313.3412227}
\bibfield{author}{\bibinfo{person}{Caojin Zhang}, \bibinfo{person}{Yicun Liu},
  \bibinfo{person}{Yuanpu Xie}, \bibinfo{person}{Sofia~Ira Ktena},
  \bibinfo{person}{Alykhan Tejani}, \bibinfo{person}{Akshay Gupta},
  \bibinfo{person}{Pranay~Kumar Myana}, \bibinfo{person}{Deepak Dilipkumar},
  \bibinfo{person}{Suvadip Paul}, \bibinfo{person}{Ikuhiro Ihara},
  \bibinfo{person}{Prasang Upadhyaya}, \bibinfo{person}{Ferenc Huszar}, {and}
  \bibinfo{person}{Wenzhe Shi}.} \bibinfo{year}{2020}\natexlab{}.
\newblock \showarticletitle{Model Size Reduction Using Frequency Based Double
  Hashing for Recommender Systems}. In \bibinfo{booktitle}{\emph{Fourteenth ACM
  Conference on Recommender Systems}} (Virtual Event, Brazil)
  \emph{(\bibinfo{series}{RecSys '20})}. \bibinfo{publisher}{Association for
  Computing Machinery}, \bibinfo{address}{New York, NY, USA},
  \bibinfo{pages}{521–526}.
\newblock
\showISBNx{9781450375832}
\urldef\tempurl%
\url{https://doi.org/10.1145/3383313.3412227}
\showDOI{\tempurl}


\end{thebibliography}

\end{document}